# Observation of Topological Surface State in High Temperature Superconductor $MgB_2$

Xiaoqing Zhou[1*], Kyle N. Gordon[1], Kyung-Hwan Jin[2], Haoxiang Li[1], Dushyant Narayan[1], Hengdi Zhao[1], Hao Zheng[1], Huaqing Huang[2], Gang Cao[1], Nikolai D. Zhigadlo[3], Feng Liu[2,4], and Daniel S. Dessau[1,5*]

[1]*Department of Physics, University of Colorado at Boulder, Boulder, CO 80309, USA*
[2]*Department of Physics, University of Utah, Salt Lake City, UT 84112, USA*
[3]*Department of Chemistry and Biochemistry, University of Bern, CH-3012 Bern, Switzerland*
[4]*Collaborative Innovation Center of Quantum Matter, Beijing, 100084, China*
[5]*Center for Experiments on Quantum Materials, University of Colorado at Boulder, Boulder, CO 80309, USA*

- Correspondence to: Xiaoqing.Zhou@Colorado.edu, Dessau@Colorado.edu

**The hunt for the benchmark topological superconductor[1,2] (TSc) has been an extremely active research subject in condensed matter research, with quite a few candidates identified or proposed. However, low transition temperatures ($T_c$) and/or strong sensitivity to disorder and dopant levels in known TSc candidates have greatly hampered progress in this field. Here, we use Angle-resolved Photoemission Spectroscopy (ARPES) to show the presence of Dirac Nodal Lines (DNLs) and the corresponding topological surface states (TSS's) on the [010] faces of the $T_c$=39K s-wave BCS superconductor $MgB_2$. Not only is this nearly triple the current record[3] of superconducting $T_c$ among all candidate TSc's, but the nature of these DNL states should make them highly tolerant against disorder and inadvertent doping variations. This makes $MgB_2$ a promising high temperature platform for the study of topological superconductivity.**

As its name suggests, a topological superconductor has two essential ingredients: nontrivial topology and the superconducting order. The exploration of topological superconductivity started with p-wave superconductors such as $Sr_2RuO_4$[4,5,6] and the 5/2 quantum Hall state in electron gas systems[7,8,9], in which the chiral superconducting order parameter is topologically nontrivial by itself. However, these p-wave superconductors are extremely sensitive to disorder, very scarce in nature, and have transition temperatures well below liquid helium temperature – each of which imposes great difficulties in the exploration of topological superconductivity. Thanks to the discovery of topological order in band structures[10,11], it was soon realized[12] that the "topological" part of a TSc can be realized with a topological surface state arising from a topologically nontrivial band-inversion, which can then be gapped by a conventional s-wave superconducting gap. This opens many new pathways to higher $T_c$ topological superconductors. One route is to construct topological insulator / superconductor hetero-structures, in which the proximity effect allows the topological surface state of a topological insulator to be gapped by the superconducting gap of a neighboring superconductor[13,14,15]. However, the proximity effect that enables topological superconductivity also imposes a severe constraint on the perfection of the interface between the two materials, which is an even greater problem for higher temperature superconductors that naturally have shorter superconducting coherence lengths. The complication of the interface physics thus becomes a major obstacle which can be naturally circumvented in a singular system, such as the one discussed here. An alternative is to dope a known bulk topological material to make it superconducting, which has seen some success in materials such as Cu-doped $Bi_2Se_3$[16,17,18]. However, given that the discoveries of high temperature superconductivity have been largely accidental, it is unclear how far this approach can be taken.

With the explosion of discoveries on numerous topological orders, there emerges another promising possibility. Since the existence of topological order in band structures seems to be more abundant than that of high temperature superconductivity, one can instead search for topological surface states in known high temperature superconductors. Indeed, it has been proposed that topological superconductivity could be found in underdoped high temperature cuprate superconductors[19] and $BaBiO_3$[20], although experimental support has not yet been reported for either of these. Very recently, a particularly promising candidate $FeTe_{1-x}Se_x$[5] was found in the family of iron-based high temperature superconductors, with a transition temperature of 14.5 K setting the current record. While this discovery is exciting, the required proximity of a very small (20 meV scale) spin-orbit-coupled gap to the Fermi energy means that the system should be highly sensitive to inadvertent doping variations. This means that an appropriate host for topological superconductivity is likely still lacking.

In this work, we use a combination of first principle density functional theory (DFT) calculations and Angle-Resolved-Photoemission Spectroscopy (ARPES) to look for topological surface states (TSS's) in $MgB_2$, a conventional BCS superconductor with a high transition temperature of 39 K[21]. Our DFT calculations[22] have predicted the existence of pairs of topological Dirac nodal lines (DNLs)[23] at the Brillouin zone boundaries as well as topological surface bands that connect these DNLs. The calculations further predict that the TSS's should be readily gapped by the superconducting order, as expected by the intimate and inherent contact between the bulk superconducting states and the topological surface states.

Although $MgB_2$ has been extensively studied by many techniques including ARPES[24, 25, 26, 27, 28], it has never been appreciated that it may harbor topological surface states. The key to the TSSs are the topological DNLs with associated Dirac points on high symmetry cuts highlighted by the arrows in the band-structure plot of Fig. 1c. As illustrated by Fig. 1d, the DNL disperses across $E_F$ in the $k_z$ direction (normal to the honeycomb layers) over a few eV range, so that Dirac band crossings as well as the corresponding TSSs will always be present at $E_F$ for essentially any conceivable amount of doping or band-bending effects. As shown in Fig. 1b, the DNLs predicted by the DFT calculation are located at the zone boundary of the 3D Brillouin zone along the K-H and K'-H' high symmetry lines that run along the z axis. Our calculations show that each of these DNL's in $MgB_2$ is wrapped by a Berry phase of π, i.e. they support a Z2 topology. Similar to the case in graphene, the K'-H' line can be regarded as the mirror image of the K-H, so the associated Berry phase for the K'-H' DNL is −π instead of π for the K-H DNL.

Because of their $k_z$ dispersion, the DNLs can be best accessed from the side so that the dispersion is in the experimental plane – a geometry different from all previous ARPES experiments that studied the sample from the c-axis, which is also the natural cleavage face. For our experiment we cleaved the samples from the "side" (blue plane of Fig 1a and 1b) and performed ARPES on that thin [010] face – a challenging but achievable task. The cleaved surface viewed with a scanning electron microscope (see Fig. 2b) as well as an atomic force microscope (see supplementary materials) shows an atomically flat region, upon which high quality ARPES spectra were observed. We measured the band dispersion along the momentum perpendicular to the cleavage face ($k_x$) by scanning the photon energy $h\nu$ from 30 eV to 138 eV along the $\Gamma$-$K$ high symmetry line. Since at these photon energies the photon momentum is negligible, we have [29]

$$k_x \approx \sqrt{\frac{2m_e}{\hbar^2}(h\nu - \emptyset - E_B + V_0) - (k_y^2 + k_z^2)} \tag{1}$$

where $\emptyset = 4.3$ eV is the work function, $E_B$ the binding energy, and $V_0 = 17\ eV$ the inner potential. This describes the ARPES spectra measured on a "spherical sheet", the radius of which is proportional to $\sqrt{h\nu - \emptyset - E_B + V_0}$. By stitching together many spectra taken over a wide range of photon energies from 30-138 eV, we reconstruct the iso-energy plots in the [001]-plane at $k_z = 0$, as shown in Fig. 2d to 2g. The data show a clear 6-fold symmetry as expected, with spectra consistent with previous ARPES studies on the [001] cleavage plane[25-29]. This confirms that we are able to observe the proper bulk band structure of $MgB_2$ from the cleaved [010] plane. As illustrated in Fig. 2c, in our experimental geometry the K-H Dirac nodal line acts as a monopole of Berry phase and is neighbored by 3 of its "mirror-image" K'-H' lines with opposite charge, each of which has bulk bands accessible only under different experimental conditions (i.e. photon energies and experimental angles). We choose the photon energy $h\nu = 86$ eV (black line in the panels), which allows us to directly access one of the K-H high symmetry lines by varying the $k_z$ momentum axis.

To better investigate the Dirac nodal line, Fig. 3 focuses on the region near the high symmetry line K-H at $k_x \approx 8\pi/\sqrt{3}a$, $k_y = 4\pi/3a$ and $k_z \sim \pi/c$ to $2\pi/c$. As shown in Fig. 3a, on the projected surface Brillouin zone we should expect a 2D topological surface state (TSS) connecting a pair of 1D Dirac nodal lines with opposite charges[30]. This "water-slide" TSS is analogous to the flat "drumhead" surface state in Dirac nodal loop systems[31], but follows the dispersions of the nodal lines over a range of ~4 eV. Viewed from the sample projection (Fig. 3b), the TSS could connect the +π nodal line ($k_y = 4\pi/3a$) to the −π nodal line on the left ($k_y = 2\pi/3a$), or the one on the right ($k_y = 8\pi/3a$), but not both. As shown in Fig. 3b to 3i, we have excellent agreement between the DFT bulk band calculations and the ARPES iso-energy plots, in which the shifting touching point of an electron pocket and a hole pocket indicates non-trivial band crossings dispersing along K-H. Importantly, there is an additional feature originating from the touching point that is absent from the DFT calculations of the bulk bands. It looks similar to a "Fermi arc" in Weyl semimetal[32], but persists for most binding energies (E-$E_F$) and connects towards the -π bulk band crossing point at $k_y = 2\pi/3a$, even though the corresponding bulk bands are not experimentally accessible at 86 eV. These look like the TSS's drawn in Fig. 2b and we label them as such, though confirmation from an energy cut (Fig. 4) is still required.

To further confirm the topological nature of the surface state, in Fig. 4 we plot the dispersion along 5 cuts across the touching points (dotted lines in Fig. 3c). As the predicted Dirac nodal line should disperse along the *K-H* high symmetry cut, the band crossings should evolve continuously from below $E_F$ to above $E_F$ as shown in Fig. 1c. This is exactly what we observed, confirming the Dirac nodal line nature of these states. We directly overlay the DFT calculations of the bulk band dispersions (white dashed lines) with the ARPES spectra with no alteration of the masses or velocities, and a -0.5 eV offset in the DFT chemical potential. The agreement is overall excellent, with the chemical potential shift indicative of extra electron charge leaving the cleaved surface. In particular, we stress that over a huge range of possible chemical shifts (such as those induced by unintentional doping), here the Dirac nodal lines dispersing along K-H guarantees Dirac points and TSS's right at $E_F$, in sharp contrast to the case of FeTe$_{1-x}$Se$_x$ (see supplementary materials for more discussions). In addition to the agreement with the bulk Dirac nodal line, the experiment shows some additional weak features. As all bulk bands are accounted for, these should be of non-bulk origin, i.e. surface states. Most interesting of these are the ones highlighted in red that are observed to connect to the Dirac points in both energy space (Fig. 4) and momentum space (Fig. 3), with this effect observed over a wide range of energies and momenta. This is fully consistent with the existence of a topological surface state (TSS) that we theoretically predicted to connect the Dirac points on this particular surface [22], and we label it as such. This state contrasts with a topologically trivial surface state (blue arrow) that is largely insensitive to the energy of the Dirac point as it disperses from cut to cut.

The observation of the topological surface state confirms the theoretical prediction of MgB$_2$ as a promising topological superconductor. Given that the realistic topological surface state in cuprate superconductors is absent, and that the transition temperature in the best pnictides superconductors are not much higher, the topological superconductivity in MgB$_2$ would not only set the current record for $T_c$ among TSc's but also approach the realistic limit. More important than the high $T_{c_2}$ however, is the fact that the topological surface state that is now expected in this material should be much less sensitive to disorder or dopant variations than in other TSc's including the newly discovered state in FeSe$_{0.45}$Te$_{0.55}$ [3]. This makes MgB$_2$ a particularly promising future platform for towards the exploration and engineering of topological superconductivity.

**Methods**

We performed first-principles calculations within the framework of density-functional theory (DFT) using the Perdew-Burke-Ernzerhof-type generalized gradient approximation (GGA) for the exchange-correlation functional, as implemented in the Vienna *ab initio* simulation package[33,34]. All the calculations are carried out using the kinetic energy cutoff of 500 eV on a 12×12×12 Monkhorst-Pack *k*-point mesh. All structures are fully optimized until the residual forces are less than 0.01 eV/Å. The electronic self-consistent iteration is converged to $10^{-5}$ eV precision of the total energy. The SOC is included in the self-consistent electronic structure calculation.

The high-quality single crystal MgB$_2$ samples have been grown by using the cubic-anvil high-pressure and high-temperature technique[35,36,37]. A mixture of Mg, B, and BN was enclosed in a BN crucible in a pyrophylite cube. A pressure of 3 GPa was applied at room temperature. Then, by keeping pressure constant, the temperature was ramped up to 1960 °C in 1 h, maintained there for 1 h, and finally decreased during 2 h. Using this method, MgB$_2$ crystals up to 0.8 × 0.8 × 0.2 mm$^3$ have been grown. For this investigation a special attention has been paid to select the samples with pronounced flat [010] surfaces. More details of the crystal growth and extensive characterization are given elsewhere[38,39,40].

Crystal structure was measured with X-ray diffraction. ARPES experiments were carried out in Stanford Synchrotron Research Laboratory (SSRL) 5-2 and Diamond Lightsource I05. All data shown in the paper were measured with 86 eV light and linear polarization. Data taken at other photon energies are available in the supplementary information.

**Data availability**

The data that support the plots within this paper and other finding of this study are available from the corresponding authors on reasonable request.

**Acknowledgements**

This work was funded by DOE project DE-FG02-03ER46066 (Colorado) and by the DOE project DE-FG02-04ER46148 (Utah). We thank Drs. D. H. Lu, M. Hashimoto, and T. Kim for technical assistance on the ARPES measurements. We thank Dr. R. Nandkishore and Dr. Qihang Liu for useful discussions. The photoemission experiments were performed at beamline 5-2 of the Stanford Synchrotron Radiation Lightsource and the Diamond Light Source beamline I05 (proposal no. SI17595). The Stanford Synchrotron Radiation Lightsource is supported by the Director, Office of Science, Office of Basic Energy Sciences, of the U.S. Department of Energy under Contract No. DE-AC02-05CH11231.

**Author Contributions**

N.D.Z. prepared and characterized the single crystals. H.D.Z, H.Z. and G.C. performed XRD analysis. K.H.J., H.Q.H. and F.L. performed DFT calculations. X.Q.Z, K.N.G., H.X.L. and D.N. performed the ARPES experiments. X.Q.Z. analyzed the data. X.Q.Z. and D.S.D. wrote the paper, while D.S.D. directed the project. All authors read and commented on the paper.

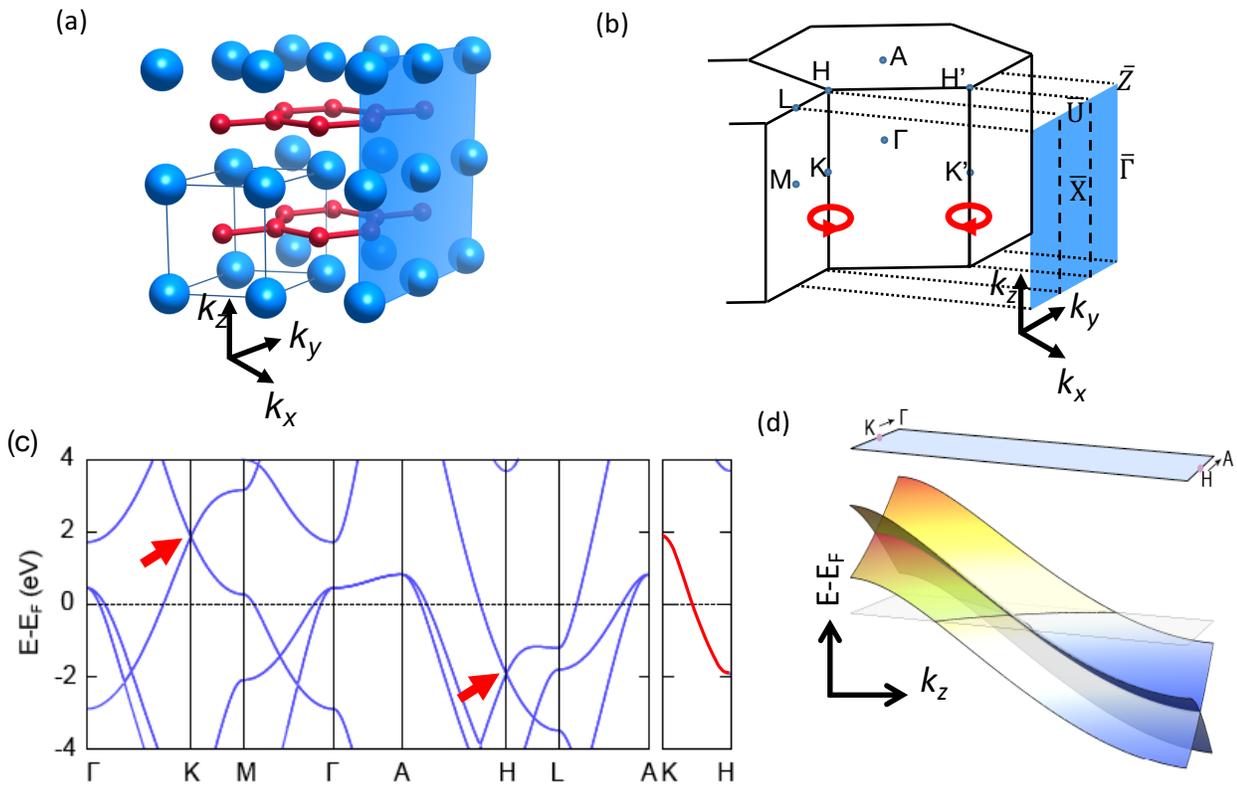

Fig. 1 a) Crystal structure of MgB$_2$ with hexagonal lattice in the ab-plane, determined by X-ray diffraction (XRD). The face of the edge-cleave is shown in blue. b) The 3D Brillouin zone and the projected "zigzag" [010] surface Brillouin zone (SBZ, shown as the blue sheet). High symmetry points K/H and K'/H' come in mirror-symmetric pairs, with Berry phase of π and –π respectively. c) Calculated DFT bulk band structure along the high symmetry cut showing Dirac points (red arrows) along K-H. d) Illustration of a Dirac nodal line along K-H, with Dirac point exactly meeting E$_F$ midway along the cut. The k$_x$ direction is normal to the [010] surface and so is covered by varying photon energy, in our case from 30-140 eV.

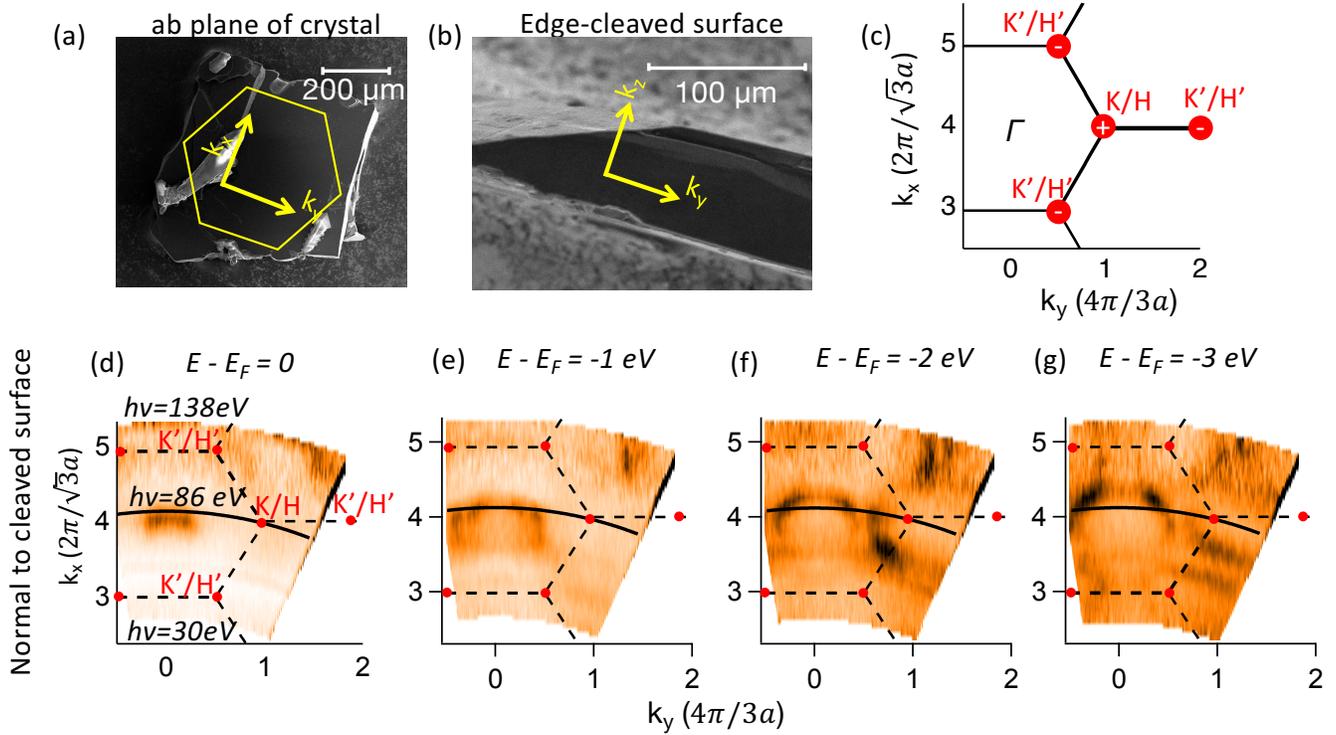

Fig 2. Edge-on ARPES gives in-plane bulk electronic structure. (a,b) In-plane and edge-cleaved views of our crystals, respectively. (c) In-plane Brillouin zone points. (d-g) In-plane isoenergy ARPES plots at energies from $E_F$ to -3 eV. The $k_y$ direction is parallel to the cleaved surface (panel b) so is covered by varying the emission angle. The $k_x$ direction is normal to the edge-cleaved surface and is covered by varying the incident photon energy from 30 eV (low $k_x$) to 138 eV (higher $k_x$). The solid line representing 86 eV cuts through the K/H Brillouin zone point.

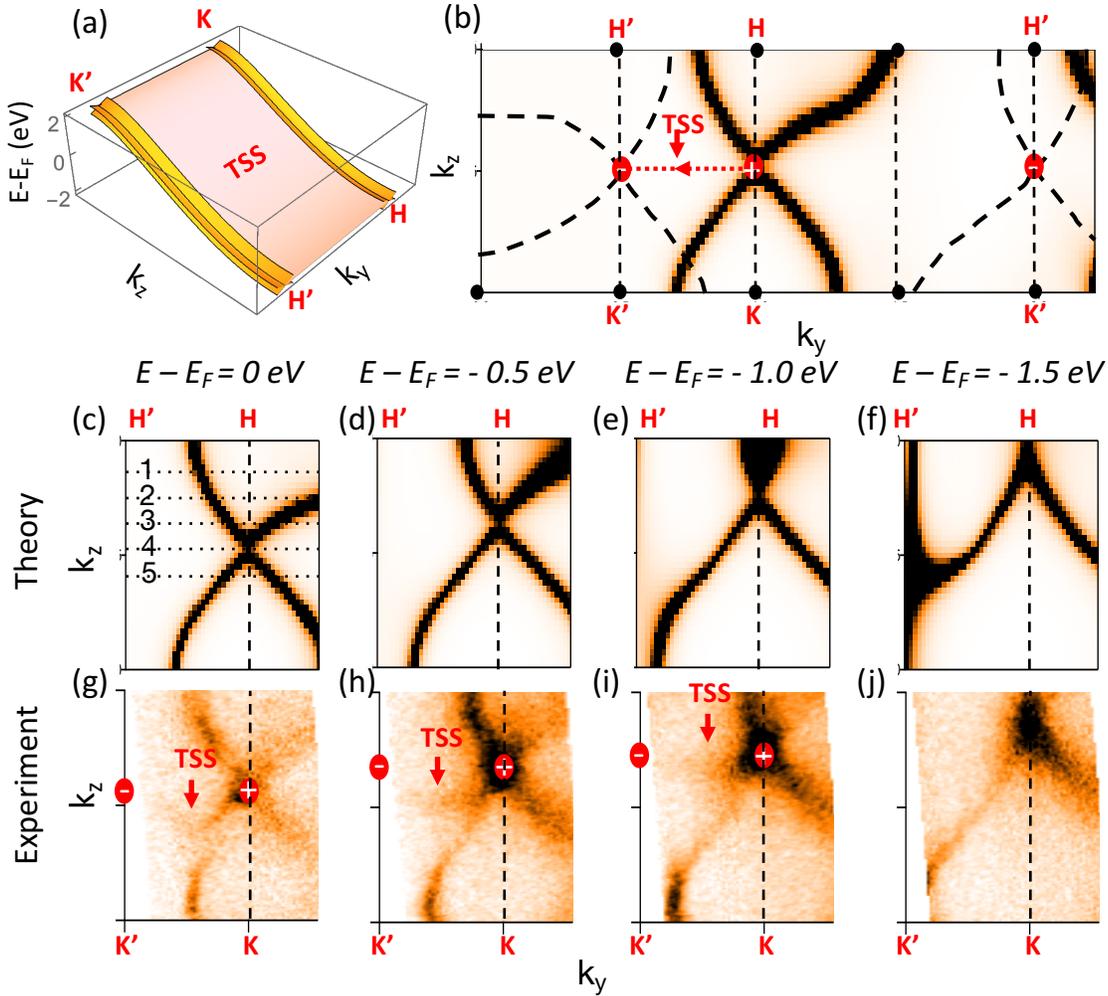

Fig 3. a) An illustration of how the 2D topological "water-slide" surface state (light magenta sheet) connects the 1D nodal lines. b) Illustration the K/H and K'/H' Dirac nodal lines with the $Z_2$ Berry phase monopoles (+ and -) projected to the 2D cleaved surface. The plot is made for the photon energy 86 eV ($k_x \approx 4$ in the units of figure 2c) that due to the proper $k_x$ value can access the bulk Fermi surface in the center of the plot. The dashed bulk Fermi surface shown at the left and right are at incorrect $k_x$ values to be observed. A topological surface state connecting the +1 and -1 monopoles on the projected surface Brillouin zone is drawn in by hand. (b-e) DFT calculation of the $k_y$-$k_z$ iso-energy contours at 86 eV at a variety of binding energies. (f-i) Experimentally measured isoenergy contours using 86 eV photons. In addition to the excellent agreement between the predicted and measured bulk states, an additional set of states laveled TSS (Topological Surface State or surface Fermi arc) conecting pairs of Dirac points are observed.

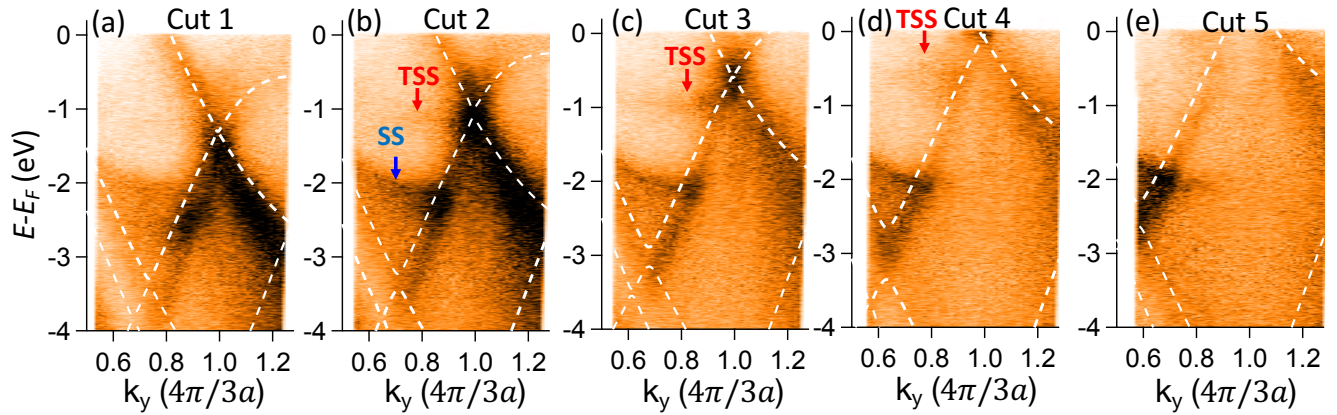

Fig. 4 Topological Surface States (TSS's) emanating from Dirac points. (a-e) ARPES spectra from the five cuts of Fig 4b, overlapped by DFT calculations (white dashed lines). The Dirac band crossings evolve from below $E_F$ to above $E_F$ consistent with the diagram of Fig 1d, with topological surface states (TSS's) originating from Dirac points indicated by the red arrows. Regular (non-topological) surface states shown by the blue arrows are also observed. The calculated DFT points have been shifted up by 0.5 eV relative to the measured spectra.

# Observation of Topological Surface State in High Temperature Superconductor MgB$_2$

## Supplementary Information

Xiaoqing Zhou[1*], Kyle N. Gordon[1], Kyung-Hwan Jin[2], Haoxiang Li[1], Dushyant Narayan[1], Hengdi Zhao[1], Hao Zheng[1], Huaqing Huang[2], Gang Cao[1], Nikolai D. Zhigadlo[3], Feng Liu[2,4], and Daniel S. Dessau[1,5*]

[1]*Department of Physics, University of Colorado at Boulder, Boulder, CO 80309, USA*
[2]*Department of Physics, University of Utah, Salt Lake City, UT 84112, USA*
[3]*Department of Chemistry and Biochemistry, University of Bern, CH-3012 Bern, Switzerland*
[4]*Collaborative Innovation Center of Quantum Matter, Beijing, 100084, China*
[5]*Center for Experiments on Quantum Materials, University of Colorado at Boulder, Boulder, CO 80309, USA*

- Correspondence to:  Xiaoqing.Zhou@Colorado.edu, Dessau@Colorado.edu

## S1. Sample and its cleaved surface

Multiple $MgB_2$ single crystals have been used for this study, with an example under the scanning electron microscope (SEM) shown in Fig. S1a. Samples are typically in the form of thin slabs, with its [001] direction perpendicular to the slab. X-ray diffraction (XRD) was used to determine the [010] direction, along which a thin line was etched into the slab to assist cleaving.

Fig. S1b shows the example of a cleaved surface under the SEM, with a uniform grayscale indicating the flatness. To further confirm the flatness of the sample, atomic force microscope was used to sample a 50 nm by 50 nm area on the cleaved surface (red dot in Fig. S1b). As shown in Fig. S1c, the surface roughness is typically below 5 nm. Therefore, not only the cleaved surface is [010], but also is atomically flat.

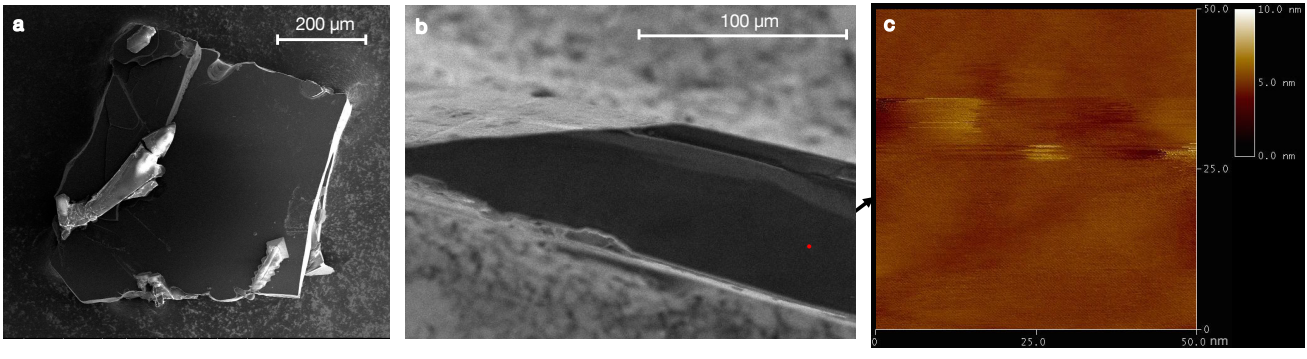

Fig. S1 a) An example of our MgB2 sample viewed from [001] direction under scanning electron microscope (SEM). b) An example of a cleaved surface on the [010] direction. c) An example of atomic force microscope.

## S2. ARPES spectra at multiple photon energies

Fig. S2 shows the comparison of ARPES iso-energy plots with the simulated DFT counterparts at photon energies 86 eV, covering the first surface Brillouin zone. At this photon energy, with the $k_z$ values available in the first zone (i.e. $k_z < \pi/c$) we can only get close to the K-H line but not exactly cover it (as shown in Fig. 1e), so the signature of the topological surface bands in the first zone is not as clear as that in the second zone, as presented in Fig. 2. On the other hand, the bulk band iso-energy plots in general agree very well with the DFT simulations.

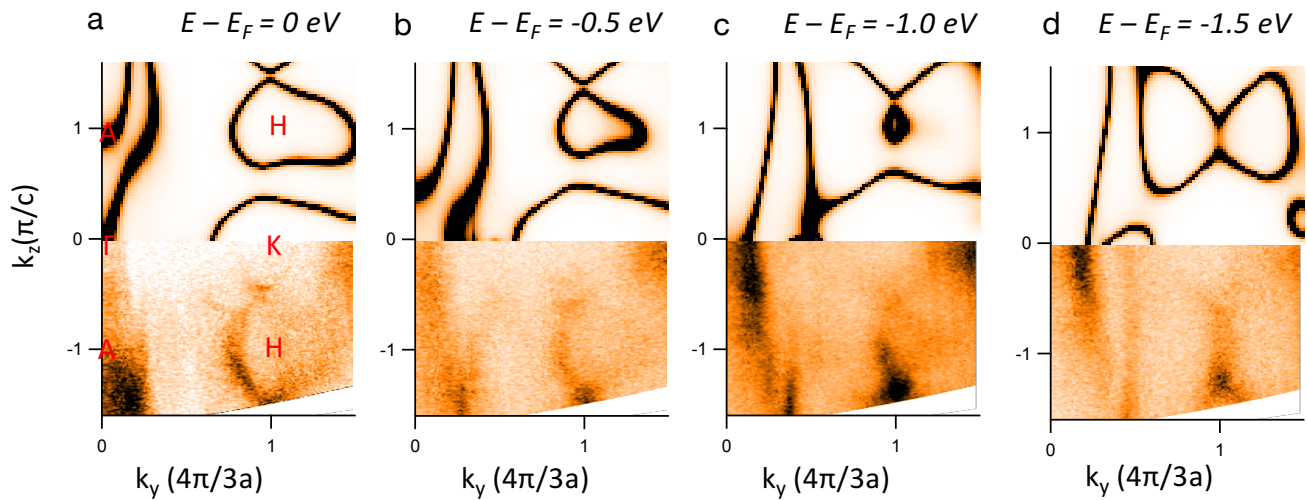

Fig. S2 Comparison of DFT simulations of the iso-enegy plot (upper half) with ARPES spectra taken at 86 eV (bottom half) at a) E-$E_F$ = 0; b) E-$E_F$ = -0.5 eV; c) E-$E_F$ = -1 eV and d) E-$E_F$ = -1.5 eV

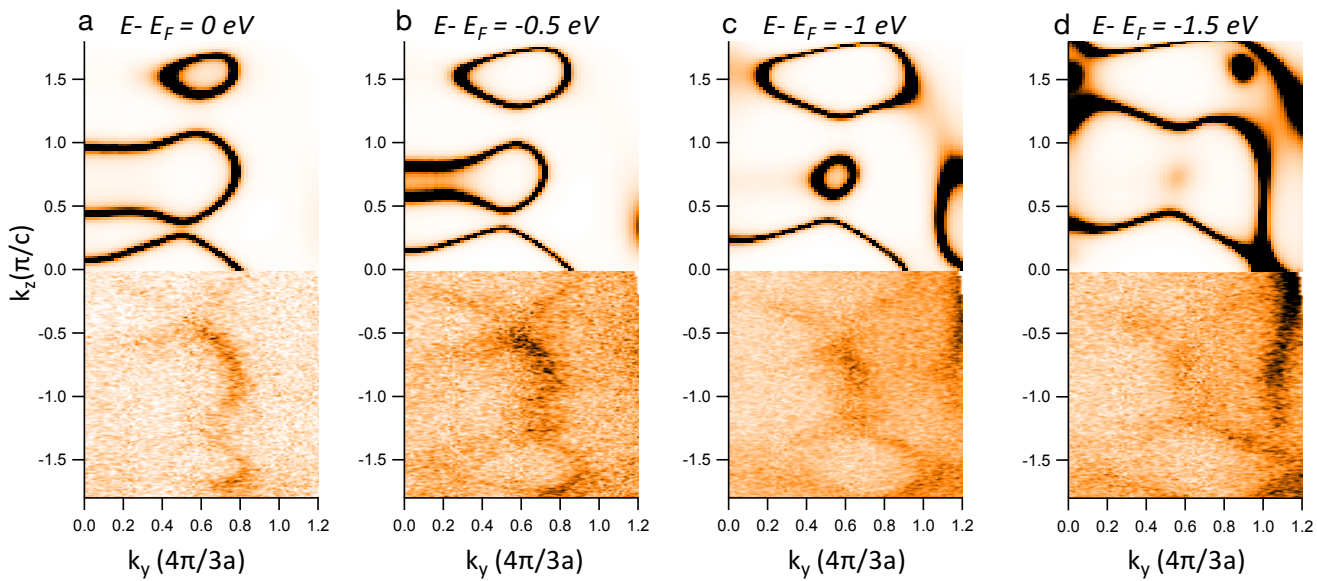

Fig. S3 Comparison of DFT simulations of the iso-enegy plot (upper half) with ARPES spectra taken at 118 eV (bottom half) at a) E-$E_F$ = 0; b) E-$E_F$ = -0.5 eV; c) E-$E_F$ = -1 eV and d) E-$E_F$ = -1.5 eV

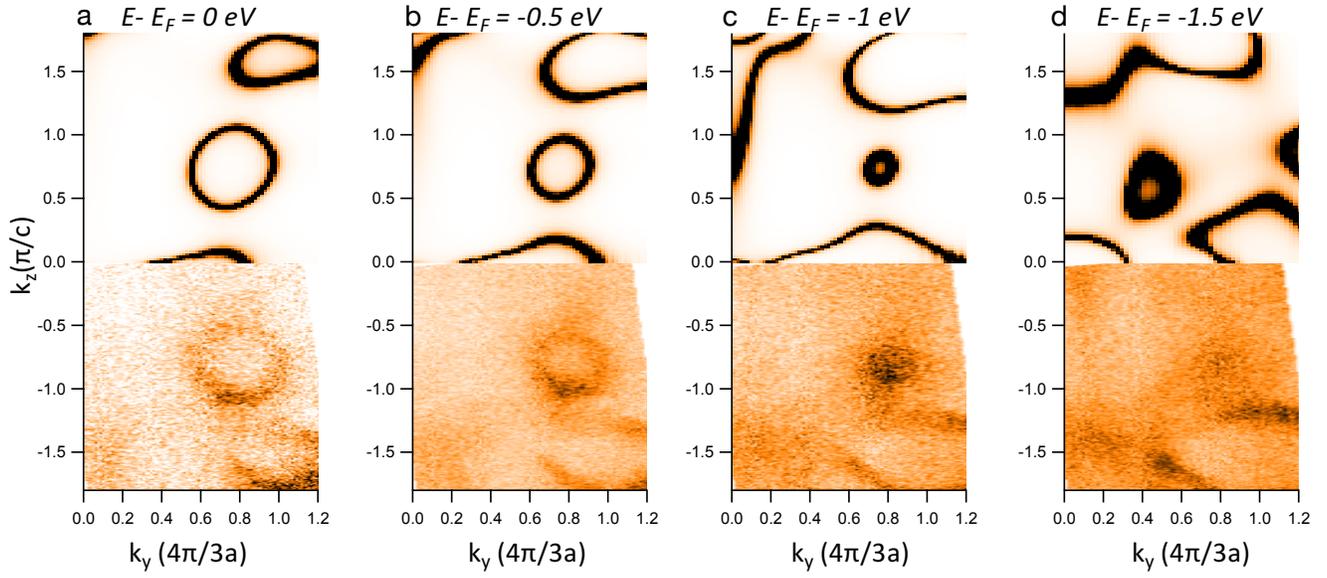

Fig. S4 Comparison of DFT simulations of the iso-enegy plot (upper half) with ARPES spectra taken at 101 eV (bottom half) at a) E-$E_F$ = 0; b) E-$E_F$ = -0.5 eV; c) E-$E_F$ = -1 eV and d) E-$E_F$ = -1.5 eV

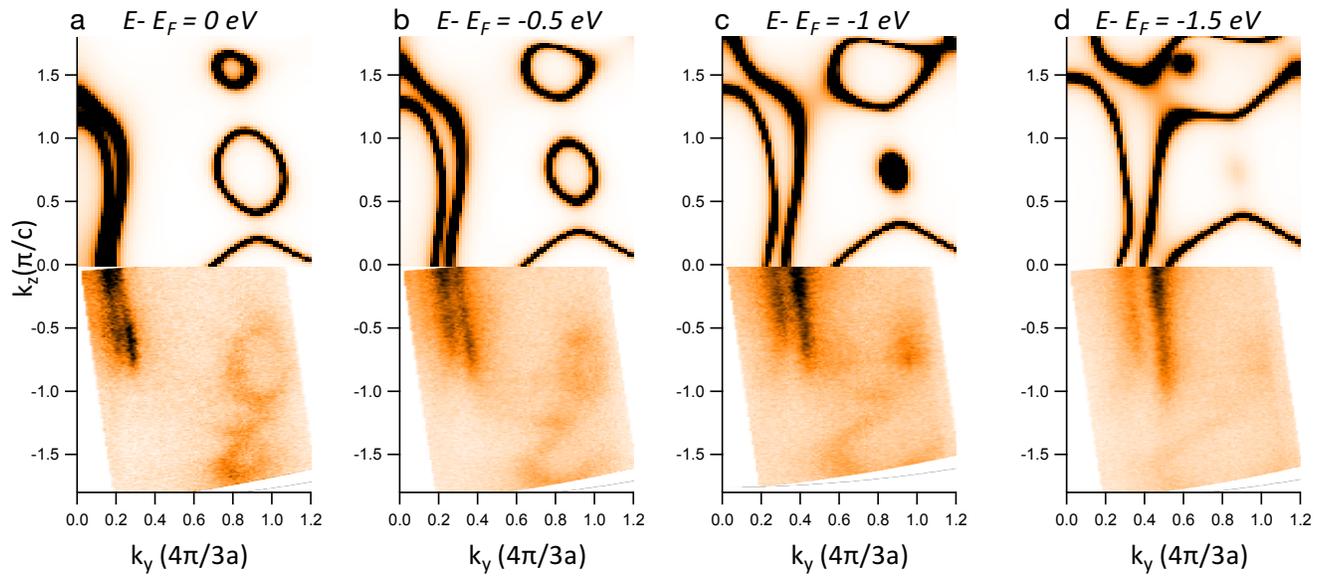

Fig. S5 Comparison of DFT simulations of the iso-enegy plot (upper half) with ARPES spectra taken at 72 eV (bottom half) at a) E-$E_F$ = 0; b) E-$E_F$ = -0.5 eV; c) E-$E_F$ = -1 eV and d) E-$E_F$ = -1.5 eV

Similarly, Fig. S3-S5 show the comparison of ARPES iso-energy plots with the simulated DFT counterparts at photon energies 118 eV, 101 eV and 72 eV respectively. Because at these photon energies the ARPES spectra would be relatively far away from the 3D K-H high symmetry line (as illustrated in Fig. 1e) at which the Dirac nodal line exists, the existence of the topological surface state is

## S3. Comparison to topological superconductivity in FeTe$_{1-x}$Se$_x$

Very recently, FeSe$_{.45}$Se$_{.55}$ was found to exhibit topological superconductivity with a transition temperature of 14.5 K, which is a record T$_c$ for single crystal TSc[1]. However, on a closer look, as shown in Fig. S6a, for FeSe$_{.45}$Se$_{.55}$ there is a very small spin-orbit coupled (SOC) gap at the crossing of the p and d bulk bands, inside of which there is a Dirac-like surface state. In ref [1] the authors showed how this Dirac like surface state exhibited a superconducting gap. Of course for this physics the p-d band crossing and the SOC must straddle E$_F$ – an extremely delicate proposition because of the small energy scale (20 meV) of the SOC gap. Any minor deviations in doping level, band bending, or related are related to move this SOC gap away from E$_F$, killing the TSc.

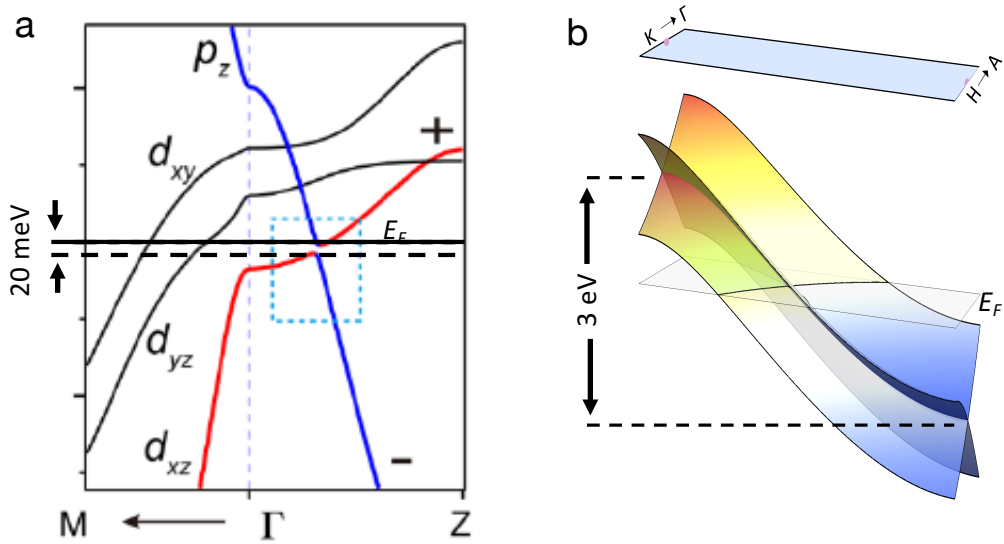

Fig. S6 Comparison of the robustness of a) FeTe$_{1-x}$Se$_x$ (reproduced from [1]) and b) MgB$_2$. For FeTe$_{1-x}$Se$_x$ the topological surface state exists within a 20 meV spin-orbital gap accidentally (by doping) at E$_F$, so it is sensitive to a small doping change. In contrast, in MgB$_2$ the Dirac nodal line disperse across 3 eV, so a Dirac band crossing can always be found at E$_F$ regardless of the unintentional doping.

In contrast, the presence of the topological surface state, is expected to be dramatically more robust in $MgB_2$. In particular, the DNL state that crosses $E_F$ has a dispersion greater than 3 eV and will thus have the relevant states precisely at $E_F$ for essentially any conceivable amount of doping or band-bending effects. There will therefore always be the appropriate states available for pairing, and the system will be highly robust. Thus the topological superconducting state should be exceedingly more robust in $MgB_2$ than in $FeSe_{.45}Se_{.55}$ – something important for the engineering of any devices based upon topological superconductivity.

---

[1] Zhang P. *et al.*, Observation of topological superconductivity on the surface of an iron-based superconductor, *Science* **360**, 182-186 (2018)